\documentclass[traditabstract]{aa} 
\usepackage{graphicx}
\usepackage{txfonts}
\usepackage[]{natbib}
\bibpunct{(}{)}{;}{a}{}{,}
\DeclareMathSymbol{\lesssim}{\mathrel}{AMSa}{"2E}

\begin{document}
\bibliographystyle{aa}
 \title{Different twins in the millisecond pulsar recycling scenario: optical polarimetry of PSR J1023+0038 and XSS J12270-4859.\thanks{Based on observations made with ESO Telescopes at the La Silla Observatory under programme ID 094.D-0692(A), (B).}}

   \author{M. C. Baglio\inst{1, 2},
          P. D'Avanzo\inst{2},
          S. Campana \inst{2},
          F. Coti Zelati \inst{1, 3, 2},
          S. Covino \inst{2}
                \and
   D. M. Russell\inst{4}          
          }
		  
   \institute{Universit\`{a} dell'Insubria, Dipartimento di Scienza e Alta Tecnologia, Via Valleggio 11, I–22100 Como, Italy                      \\
              \email{cristina.baglio@brera.inaf.it}
         \
             \and 
             INAF, Osservatorio Astronomico di Brera, Via E. Bianchi 46, I-23807 Merate (LC), Italy
             \and
             Anton Pannekoek Institute for Astronomy, University of Amsterdam, Postbus 94249,  NL-1090-GE Amsterdam, The Netherlands
           \and
            New York University Abu Dhabi, P.O. Box 129188, Abu Dhabi, United Arab Emirates             
              }          
   \date{ }

   \abstract{
   We present the first optical polarimetric study of the two transitional pulsars PSR J1023+0038 and XSS J12270-4859. This work is focused on the search for intrinsically linearly polarised optical emission from the two systems.
   To this aim, we carried out multiband optical ($BVRi$) and near--infrared ($JHK$) photo--polarimetric observations of the two systems using the ESO New Technology Telescope (NTT) at La Silla (Chile), equipped with the EFOSC2 and the SOFI instruments.
   The system XSS J12270-4859 was observed during its radio--pulsar state; we did not detect a significant polarisation degrees in all bands, with $ 3\sigma $ upper limits of, e.g., 1.4$\%$ in the $R$--band. We built the NIR--optical averaged spectral energy distribution (SED) of the system, that could be well described by an irradiated black body with radius $ R_{*} = 0.33\pm0.03\, R_{\odot}$ and albedo $ \eta=0.32\pm 0.05 $, without the need of further components (thus excluding the visible presence of an extended accretion disc and/or of relativistic jets).
   The case was different for PSR J1023+0038, that was in its accretion phase during our campaign. We measured a linear polarisation of $1.09\pm0.27\%$ and $0.90\pm 0.17\%$ in the $ V $ and $ R $ bands, respectively. The phase--resolved polarimetric curve of the source in the $ R $-- band reveals a hint of a sinusoidal modulation at the source 4.75 hr orbital period, peaked at the same orbital phase as the light--curve.
The measured optical polarisation of PSR J1023+0038 could in principle be interpreted as electron scattering with free electrons (that can be found in the accretion disc of the system or even in the hot corona that sorrounds the disc itself) or to synchrotron emission from a relativistic particles jet or outflow. However, the NIR--optical SED of the system built starting from our dataset did not suggest the presence of a jet. We conclude that the optical linear polarisation observed for PSR J1023+0038 is possibly due to Thomson scattering with electrons in the disc, as also suggested from the possible modulation of the $ R $--band linear polarisation at the system orbital period. }

\authorrunning{Baglio et al.} 
\titlerunning{Optical polarimetry of PSR J1023+0038 and XSS J12270-4859}
\maketitle

\section{Introduction}
Millisecond radio pulsars (MSPs) have been believed for years to be the descendants of old, weakly magnetized neutron stars (NSs) that are hosted in low mass X--ray binaries (LMXBs). The transfer of angular momentum from the low-mass companion to the NS through accretion is in fact considered as the possible responsible of the spin--up of slow NSs in ms spinning sources (\citealt{Alpar82}; \citealt{Radhakrishnan82}). The initial NS, after switching off its radio pulsations while entering the so called \textit{pulsar graveyard}, might be re-accelerated through accretion and switch on again as a recycled MSP once the mass tranfer is over (this is the so called \textit{recycling scenario} of MSPs; \citealt{Srinivasan2010}). During the accretion phase, X--ray pulsations at the spin frequency might be detected. In this case, the system takes on the typical features of an accreting millisecond X--ray pulsar (AMXP). Fifteen AMXPs have been discovered so far \citep{PatrunoReview}.

The recycling scenario of MSPs discussed above was initially confirmed by the so called ``missing link pulsar'', PSR J1023+0038 \citep{Archibald2009}, which was the first source showing the potential to alternate a radio pulsar phase to an X--ray state, powered by accretion. 
Because of the repeated transitions that systems like PSR J1023+0038 undergo from an accretion state to a rotation powered state, they are often referred to as transitional MSPs.

After decades of search however, the first observational direct evidence of the link between MSPs and NS-LMXBs has been found thanks to the detection of transient millisecond X--ray pulsations during a few weeks--long X--ray outburst from the transient LMXB IGR J18245--2452 \citep{Papitto2013}, followed by the turn on of a radio MSP, a few weeks later. A similar system is the LMXB XSS J12270--4859: radio, optical and X-–ray observations over the last years suggested that it may have switched to a MSP phase since the end of 2012 (\citealt{Bassa2014}; \citealt{Bogdanov2014}; \citealt{deMartino2015}), and the detection of 1.69--ms radio pulsations provided compelling evidence for the state change \citep{Roy2014}.

Optical and infrared observations of such systems are frequently performed, with both imaging and spectroscopic techniques.  
Most of the emission at these wavelengths is expected to originate from the accretion disc that surrounds the NS or from the companion star. In particular, observations during quiescence offer a unique possibility to study the properties and characteristics of the companion star. Less used, but equally interesting, are optical and infrared linear polarimetric studies of LMXBs. The emission from these objects may in fact be polarised due to different reasons, some of them intrinsic to the sources. A first possible cause of linear polarisation lies in the accretion disc of LMXBs, that are heated from the compact object X--ray emission, causing the hydrogen to be almost completely ionised. In these conditions electron scattering of radiation (initially emitted unpolarised) can occur in the geometrically symmetrical accretion disc, resulting in linearly polarised radiation at a constant level. Any phase--dependent variations may be due to asymmetries in the disc geometry or in the system configuration.
However, each electron that is responsible for the scattering in the disc oscillates in principle in a random direction. Therefore the large amount of expected linear polarisation caused by electron scattering never exceeds a few per cent in the optical and infrared bands for LMXBs, due to strong cancellation effects. An alternative and intriguing scenario links the detection of linear polarisation in LMXBs to the possible emission of a relativistic particle jet, similarly to what happens in active galactic nuclei. The radiation emitted from a leptonic jet has in fact a synchrotron nature, which is known to be intrinsically linearly polarised up to tens of per cent. Even in this case, due to the presence of strongly tangled magnetic fields in the jet, radiation is expected to be polarised at a few per cent at most, mainly at lower frequencies (the radio and infrared bands; \citealt{Russell08}).



\section{Targets of this work}
\subsection{PSR J1023+0038}
The first system that showed the potential to alternate its radio pulsar phase, powered by rotation, to an X--ray phase, powered by accretion, was the 1.69--ms radio pulsar in a 4.75 hr binary orbit PSR J1023+0038 (hereafter J1023). This system was discovered by \citet{Bond2002} in the radio band and initially classified as a magnetic cataclysmic variable. Optical studies revealed signs for the presence of an accretion disc in 2001 \citep{Szkody2003}, which led to identify it as a NS--LMXB. No enhanced X--ray emission was reported. 
In 2007 an 1.69--ms radio pulsar was discovered to be the compact object of the system \citep{Archibald2013}. J1023 underwent a state change from the MSP state to an accretion--disc dominated state in June 2013, with a consequent switch off of the radio pulsed signal and a significant increase of the GeV flux \citep{Stappers2014}.

The system has been the subject of extensive multi--wavelength campaigns after its state change (\citealt{Halpern2013}; \citealt{Kong2013}; \citealt{Linares2014}; \citealt{Patruno2014}; \citealt{Takata2014}; \citealt{Tendulkar2014}; \citealt{CotiZelati2014}; \citealt{Deller2015}; \citealt{Archibald2015}; \citealt{Bogdanov2015}; \citealt{Shahbaz2015}). Radio band observations performed by \citet{Deller2015} in particular showed a rapidly variable and flat spectrum persisting over six months.

During the multi--wavelength observational campaign reported in \citet{Bogdanov2015}, the system was observed to exhibit short-lived but frequent and stable episodes of mode switching in the X--rays (factor of $\sim 6$ in flux) occurring on timescales of $ \sim 10 $s. 
In the UV and in the optical, \citet{Bogdanov2015} reported on occasional intense flares that were coincident with those observed in the X--rays; radio observations revealed no pulsations at the pulsar period during any of the three X--ray modes, probably due to the quenching/screening of the radio emission mechanisms by the accretion flow.
\citet{Shahbaz2015} reported on an optical campaign performed on J1023 with the WHT and the TNT telescopes. The highest time-resolution (0.31 and 0.97 or 3.37 s respectively) light curves showed for the first time the presence of rectangular dip features randomly distributed with the orbital phase, which are similar to the mode-switching behaviour observed in the X--ray light curves of the source.

The companion star of the system has been spectroscopically classified as a $G$--type star (moderately irradiated by the compact object) and has a mass of $ \sim 0.2\, M_{\odot} $ \citep{Archibald2009}. The distance of the system is 1.37 kpc \citep{Deller2012}, from which the authors derived an estimate of the mass of the pulsar ($1.71 \pm 0.16 M_{\odot} $). Finally \cite{Archibald2013} evaluated the system inclination as $ 42^{\circ}\pm2^{\circ} $.

\subsection{XSS J12270-4859}

Discovered by the Rossi X-ray Timing Explorer \citep{Sazonov2004}, XSS J12270--4859 (hereafter J12270) was initially classified as a cataclysmic variable hosting a magnetic white dwarf (\citealt{Masetti2006}; \citealt{Butters2008}), based on the presence of optical emission lines. Subsequent independent multi-wavelength studies (\citealt{Pretorius2009}; \citealt{Saitou2009}; \citealt{deMartino2010}; \citealt{Hill2011}; \citealt{deMartino2013}; \citealt{Papitto2014}) casted doubts on this classification, suggesting instead an identification as a LMXB showing unusual dipping and flaring behaviour on time scales of few hundreds of seconds. XSS J12270--4859 was subsequently recognized to be spatially coincident with a moderately bright gamma-ray source detected by Fermi-LAT emitting up
to 10 GeV (\citealt{deMartino2013} and references therein), and now known as 3FGL J1227.9--4854.

The system has remained stable in gamma-rays, X-rays and optical for about a decade until 2012 November/December, when a substantial decline in brightness in all these bands was reported
(\citealt{Bassa2013}; \citealt{Tam2013}; \citealt{Bassa2014}; \citealt{Bogdanov2014}). This unexpected variability led to the prediction that the system could harbour an active radio MSP, sharing similar properties with J1023. Follow-up observations in the radio finally succeeded in detecting pulsations, at a period of 1.69 ms \citep{Roy2014}.

The orbital period of the system is 6.91 hr (\citealt{Bassa2014}; \citealt{deMartino2014}), and the companion is a G5-type, donor star with mass of 0.15--0.36 M$_{\odot}$ \citep{deMartino2014}, which dominates the UV/optical and near-IR emissions. The distance is estimated to be about 1.4 kpc, and the analysis of optical light curves constrains the binary inclination in the $46^\circ \lesssim i \lesssim 65^\circ$ interval \citep{deMartino2015}.


\section{Optical polarimetry with EFOSC2}\label{Opt_pola}
The systems J1023 and J12270 were observed in quiescence on 8 and 9 February 2015 (respectively) with the ESO New Technology Telescope (NTT) located at La Silla (Chile), equipped with the EFOSC2 camera in polarimetric mode, using the $B$, $V$, $R$, $i$ filters.
The night was photometric, with seeing constant at a level of $\sim 0.6-0.8 ''$ , and the Moon at the 84$\%$. 
We performed image reduction following the standard procedure of subtraction of an average bias frame and division by a normalised flat frame. We performed the flux measurements by using aperture photometry techniques on the fields of the two targets with {\tt daophot} \citep{Stetson1987}.
In order to gain polarimetric observations, a Wollaston prism was inserted in the path of radiation. In this way, the incident radiation is split into two beams with orthogonal polarisation, called ordinary ($o-$) and extraordinary ($e-$) beams. A mask allowed the different beams not to overlap on the CCD.
The instrument was also equipped with a rotating half wave plate (HWP), that is able to rotate the polarisation plane of the incident radiation in a direction that is symmetrical to its optical axis. In particular, the linear polarisation direction of the radiation exiting the plate rotates at a double velocity with respect to the angular velocity of the plate itself.

The linear polarisation (LP) of radiation is entirely described by the Stokes parameters $ Q $ and $ U $, defined as follows:
\begin{equation}\label{st_par}
Q=f^{o}(0^{\circ})-f^{e}(0^{\circ});\\
U=f^{o}(45^{\circ})-f^{e}(45^{\circ}),
\end{equation} 
where $f^{o}(0)$, $f^{e}(0)$, $f(^{o}45)$, $f^{e}(45)$ are the ordinary ($o$) and extraordinary ($e$) fluxes of the linearly polarised components at $0^{\circ}$ and $45^{\circ}$ degrees, respectively (angles are intended with respect to the telescope axis). In order to obtain an estimate of $ Q $ and $ U $ it would thus be enough to take two images (with the Wollaston prism mounted on the instrument) at the position angles $ \Phi = 0^{\circ}$ and $45^{\circ}$, e.g. by rotating the whole instrument by the two angles with respect to the telescope axis. However, to increase the precision of the polarimetric measurement, images at four different angles are often taken (namely $ \Phi = 0^{\circ}, 45^{\circ}, 90^{\circ}, 135^{\circ}$). Since EFOSC2 is equipped with a HWP that can rotate, it is also possible to avoid rotating the whole instrument; for this reason, images were taken at the four angles of the HWP rotator $ \Phi_i=22.5^{\circ}(i-1), i=1, 2, 3, 4 $ (that correspond to a rotation of the instrument by $ \Phi = 0^{\circ}, 45^{\circ}, 90^{\circ}, 135^{\circ}$, respectively). 


Alternating the filters, we obtained for the targets J1023 and J12270 respectively a set of five 60s integration images and four 120s integration images, for each filter and HWP angle (a log of the optical observations is reported in Tab. \ref{log_optical}).

    \begin{table}
\caption{Complete log of the NTT optical observations performed on 2015 February, 8.}             
\label{log_optical}      
\centering                          
\begin{tabular}{c c c c}        
\hline\hline                 
Target & Filter  & Exposure time& UT mid observation  \\    
        & &  (per pos angle)                 & (YYYmmdd)\\
\hline                        
J1023 & $B$ &$5\times60$ s & 20150208.2898\\
                           & $V$ &$5\times60$ s & 20150208.2944\\
                           & $R$ &$5\times60$ s &20150208.2990\\
                           & $i$ &$5\times60$ s  &20150208.3036\\
\hline                                   
J12270 & $B$ &$4\times120$ s & 20150209.2110\\
                           & $V$ &$4\times120$ s & 20150209.2184\\
                           & $R$ &$4\times120$ s &20150209.2257\\
                           & $i$ &$4\times120$ s  &20150209.2330\\
\hline
\end{tabular}
\end{table}

In this configuration, eq. \ref{st_par} for deriving $ Q $ and $ U $ translates into the following equations for the normalised Stokes parameters (a precise derivation is found in \citealt{Serego}, \citealt{Tinbergen1996} and \citealt{Schmid08}):

\begin{equation}\label{stokes}
Q=\frac{F(\Phi_{1})-F(\Phi_{3})}{2} ; \,\,\,\,\, U=\frac{F(\Phi_{2})-F(\Phi_{4})}{2},
\end{equation}
where
\begin{equation}
F(\Phi_{i})=\frac{f^{o}(\Phi_{i})-f^{e}(\Phi_{i})}{f^{o}(\Phi_{i})+f^{e}(\Phi_{i})},
\end{equation}
and $ f^{o}(\Phi_{i}) $, $ f^{e}(\Phi_{i}) $ are the ordinary ($o$) and extraordinary ($e$) fluxes observed with the HWP rotator oriented at $\Phi_{i}=22.5^{\circ}(i-1), i=1, 2, 3, 4$. 

Once the Stokes parameters are worked out, it is possible to obtain an estimate of the LP degree simply by computing the quantity:
\begin{equation}\label{pol_deg}
P=\sqrt{Q^2+U^2}.
\end{equation}

This simple expression however does not take into account neither the possible LP induced by the interaction of the radiation emitted by our targets with the interstellar dust along the line of sight, nor any possible instrumental effect that could induce the polarisation of the light (e.g. the reflection on the $45^{\circ}$ mirror of the telescope, that introduces strong instrumental effects when the telescope has a Nasmyth focus, as in the case of the NTT). 
For low polarisation levels one possibility is to correct the values of $ Q $ and $ U $ of the targets obtained as described above by using the Stokes parameters of an unpolarised standard star (or the weighted mean of the Stokes parameters evaluated for a group of field stars, supposed to be unpolarised). In addition, since the Stokes parameters statistics is not Gaussian (\citealt{Wardle1974}; \citealt{Serego}), the LP degree obtained with eq. \ref{pol_deg} after the correction of $ Q $ and $ U $ must be in turn corrected for a bias factor as described by the following equation:
\begin{equation}\label{pol_bias}
P=P_{\rm obs}\sqrt{1-(\frac{\sigma_{\rm P}}{P_{\rm obs}})^2},
\end{equation}
where $ P_{\rm obs} $ and $  \sigma_{\rm P}$ are the LP degree obtained by using eq. \ref{pol_deg} and the rms error on the LP degree, respectively.

Alternatively, one could compute, for each value of the instrumental position angle $ \Phi $, the quantity:
\begin{equation}\label{eq_S_par}
S(\Phi)=\left(\frac{f^{o}(\Phi)/f^{e}(\Phi)}{\left\langle f^{o}_{u}(\Phi)/f^{e}_{u}(\Phi_{i})\right\rangle  }-1\right)/\left(\frac{f^{o}(\Phi)/f^{e}(\Phi)}{\left\langle f^{o}_{u}(\Phi)/f^{e}_{u}(\Phi)\right\rangle  }+1\right),
\end{equation}
where $ \left\langle f^{o}_{u}(\Phi)/f^{e}_{u}(\Phi)\right\rangle  $ is the averaged ratio between the ordinary and extraordinary fluxes of the unpolarised field stars (chosen as reference), that are supposed to be subject to the same extinction as the target itself.
As exhaustively described in \citet{Serego}, this parameter gives an estimate of the projection of the LP along the different directions. 
The parameters $ S $ and the LP degree and angle ($ P $ and $ \theta_{\rm P} $, respectively) are then related through the relation:

\begin{equation}\label{fit_cos}
S(\Phi)= A+ P\cos 2\left( \theta_{\rm P} -\Phi\right),
\end{equation}

where $ A $ is a constant.
The fit of $ S(\Phi) $ with eq. \ref{fit_cos} yields $ P $ from the semi-amplitude of the oscillation and $\theta_{\rm P}$ from the position of the first maximum of the curve. From the constant $ A $, indeed, the presence of a residual interstellar or instrumental component to the LP is proven.

The $ S $ parameter is already normalised to the non-polarised reference stars; in this way, it is guaranteed to obtain values of $ P $ that are automatically corrected both for interstellar and instrumental effects. In case any residual effect is present, it would be simply indicated by the observation of a shift with respect to 0 along the $ y- $ axis of the sinusoidal function reported in eq. \ref{fit_cos}. Moreover, with this method no bias correction (eq. \ref{pol_bias}) due to the non--Gaussianity of the polarisation measurement is needed if proper techniques to derive confidence intervals in multi-parametric fits are applied.



\section{NIR polarimetry with SOFI}\label{NIR_pola}
The systems J1023 and J12270 were observed on 2015, February 7 in photo-polarimetric mode with the NTT equipped with the SOFI instrument in the near--infrared ($ J $, $ H $, $ K $ bands). The night was clear, with seeing $\sim 0.8''$ during the whole night. In order to obtain a precise illumination correction\footnote{\url{https://www.eso.org/sci/facilities/lasilla/instruments/sofi/tools/reduction/flat_fielding.html}} of our frames, we performed image reduction by subtracting an average dark frame and dividing by a normalised special flat frame, obtained through a {\tt midas} procedure available on the ESO website\footnote{\url{https://www.eso.org/sci/facilities/lasilla/instruments/sofi/tools/sofi_scripts.html}}. All the flux measurements have been performed with the {\tt daophot} task, as for the optical (Sec. \ref{Opt_pola}). Even in SOFI, a Wollaston prism guarantees the split of incident radiation in the two orthogonally polarised $o-$ and $e-$ beams. As in the case of EFOSC2, a mask prevents any overlapping of the two beams. Nevertheless in this instrument a rotating HWP is not provided; for this reason, in order to obtain the images at different angles needed to perform a LP measurement it is necessary to rotate cyclically the whole instrument with respect to the telescope axis of an angle $ \phi_{i}=45^{\circ}(i-1) $, with $ i= 1, 2, 3, 4 $. In this way, the observed field appears always different between an image and its consecutive, that is rotated of $ 45^{\circ} $ with respect to the previous one. For this fact, unless the observed field is particularly crowded (that is not the case of our two targets), it is difficult to obtain in the infrared a group of (or even one) field stars that can be imaged for each orientation of the instrument and that can be used as reference unpolarised stars to correct the Stokes parameters of the target for interstellar and instrumental effects.
Moreover, due to the Nasmyth focus of the instrument, the entity of these effects depends on the direction of the observation, and for this reason one needs to observe objects that are exactly at the same azimuth of the targets in order to obtain a precise correction (this would be the case of the reference field stars, if present), or to implement a model that is able to take into account all the possible instrumental contaminations to the polarimetric measurements (similarly to what already done for the PAOLO instrument mounted at the Telescopio Nazionale Galileo -- TNG; \citealt{Covino_PAOLO}). This model is at the moment under construction, and the SOFI polarimetric images remain thus still uncalibrated.

\section{Results}
\subsection{NIR and optical imaging of J1023 and J12270}\label{opt_ima}
\begin{figure}
\centering
\includegraphics[scale=0.44]{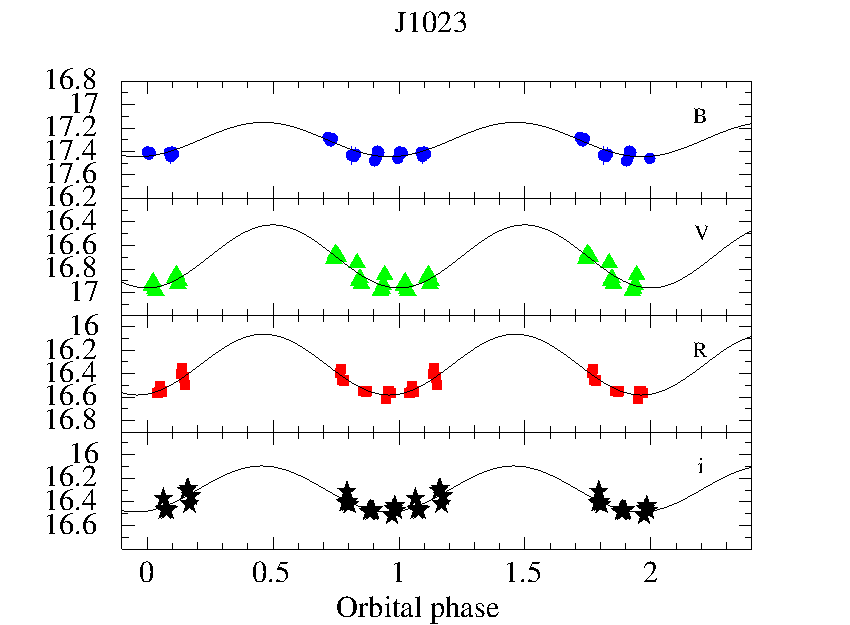}
\caption{From top to bottom, $BVRi$ light curves (magnitudes vs. orbital phase) of J1023. Magnitudes are not corrected for reddening. Superimposed, the sinusoidal fits of the curves with period fixed to 1. Errors are indicated at the 68\% confidence level. Two periods are drawn for clarity.}
\label{lightcurves_J1023}
\end{figure}

\begin{figure}
\centering
\includegraphics[scale=0.43]{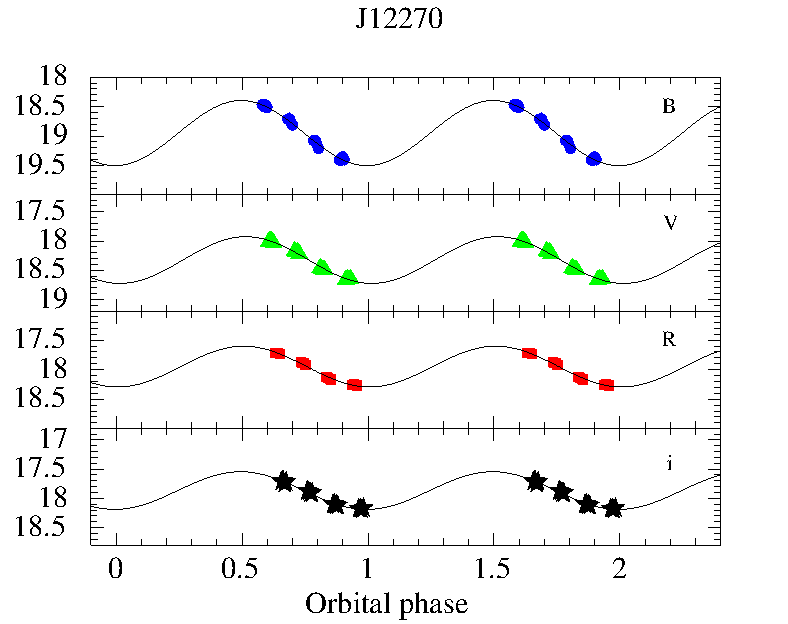}
\caption{From top to bottom, $BVRi$ light curves (magnitudes vs. orbital phase) of J12270. Magnitudes are not corrected for reddening. Superimposed, the sinusoidal fits of the curves with period fixed to 1. Errors are indicated at the 68\% confidence level. Two periods are drawn for clarity.}
\label{lightcurves_J12270}
\end{figure}

The photo--polarimetric images of J1023 and J12270 allowed us to extract their optical--NIR fluxes. While in the NIR we just summed together all the images in order to enhance the signal to noise ratio of the observations, in the case of the optical images we could build the $ BVRi $ light curves of the system. We computed the total flux of each star in the two fields simply by summing the intensities of the ordinary and extraordinary beams in all the collected images (see Sec. \ref{Opt_pola} and \ref{NIR_pola} for details on the flux measurements). We then performed differential photometry with respect to a selection of isolated field stars, with the aim of minimising any systematic effect. Finally we calibrated the infrared magnitudes using the 2MASS\footnote{\url{http://www.ipac.caltech.edu/2mass/}} catalog, the optical magnitudes of J1023 using the SDSS\footnote{\url{http://www.sdss3.org/dr10}} catalog and the optical magnitudes of J12270 using the APASS\footnote{\url{https://www.aavso.org/download-apass-data}} catalog (we used the transformation equations of \citealt{Jordi2006} to pass from the SDSS and APASS $ gri $ magnitudes to the $ BVR $ Johnson--Cousin ones).
The optical light curves of the two sources can all be fitted with sinusoidal functions that are modulated at the known orbital periods of the systems (4.75 hr and 6.9 hr for J1023 and J12270, respectively). The results of the NIR and optical photometry of J1023 and J12270, together with the results of the sinusoidal fits in the optical, are reported in Tab. \ref{Tab_NIRopt_phot} and are shown in Fig. \ref{lightcurves_J1023} and \ref{lightcurves_J12270}. 

\begin{table*}[htb]
\caption{Results of the NIR and optical photometry performed on J1023 and J12270. The magnitude values (Vega magnitudes for the $ JHKBVR $ filters; AB magnitudes for the $ i $ filter) are not corrected for reddening, the parameters of which are reported in the last 
column (the $ A_{V} $ has been derived from \citet{Schlafly2011} for J12270 and from the $ N_{\rm H} =5.3\cdot10^{20} \, \rm cm^{-2}$ estimate reported in Campana et al. in preparation for J1023). In the $\rm 6^{\rm th}$ coloumn, the reduced $\chi^2$ of the sinusoidal fits of the optical light curves are reported.}
\label{Tab_NIRopt_phot}      
\centering                          
\begin{tabular}{c c c c c c c c}       
\hline\hline                 
Filter   & Frequency & Semi-Amplitude & Maximum &Mean  & $\chi^2/\rm dof$&$A_{\lambda}$ \\   
         &       (Hz)       & (mag)    & (phase)                 & magnitude & &(mag) 	 \\
\hline
\multicolumn{7}{|c|}{J1023}\\
\hline                       
   $J$ &  $ 2.41\times10^{14} $ &-- &-- &$15.23 \pm 0.03$& --&$0.076 \pm 0.009 $\\
   $H$ & $ 1.81\times10^{14} $ &-- &-- &$ 14.98 \pm 0.04 $ &  --&$0.049 \pm 0.006$ \\
   $K$ &$ 1.39\times10^{14} $  &-- &--  &$14.45 \pm 0.06 $& --&$0.033\pm 0.004$ \\%
    $i$ & $ 3.78\times10^{14} $  &$ 0.19\pm 0.02 $ &$ 0.45\pm0.01 $& $16.29\pm 0.02$ & 120.00/17& $ 0.17\pm 0.02$\\
   $R$ & $ 4.67\times10^{14} $  &$0.26\pm 0.02$&$0.46\pm 0.01$& $16.32\pm 0.01$ & 131.80/17&$ 0.23\pm 0.03$\\
   $V$ & $ 5.48\times10^{14} $  &$0.27 \pm 0.01$ &$0.50\pm0.01$& $16.69\pm 0.01$ & 232.20/17& $ 0.29\pm 0.03$\\
   $B$ & $ 6.82\times10^{14} $  &$ 0.15\pm 0.03$ &$ 0.46\pm 0.02 $& $17.30 \pm 0.03$ & 27.46/17&$ 0.38\pm 0.05$\\
\hline                                  
\multicolumn{7}{|c|}{J12270}\\
\hline   
  $J$ & $ 2.41\times10^{14} $  &-- & --&$16.94 \pm 0.06$&--  &$0.093 \pm 0.003 $\\
   $H$ & $ 1.81\times10^{14} $  &-- &-- &$ 16.63 \pm 0.11 $  &-- &$0.060 \pm 0.002$ \\
   $K$ & $ 1.39\times10^{14} $   &-- &-- &$16.21 \pm 0.21 $ &-- &$0.041\pm 0.001 $ \\%
       $i$ & $ 3.78\times10^{14} $  &$0.32\pm 0.08$ & $0.49\pm0.06$&$17.87\pm 0.15$  & 0.58/13&$ 0.20\pm 0.01$\\
   $R$ & $ 4.67\times10^{14} $  &$0.34\pm 0.03$ &$0.51\pm0.03$ &$17.95\pm 0.05$ & 5.00/13 &$ 0.28\pm 0.01$\\
   $V$ & $ 4.66\times10^{14} $  &$0.40\pm0.02$ &$0.48\pm0.03$ &$18.32\pm 0.05$ & 5.00/13 &$ 0.36\pm 0.01$\\
   $B$ & $ 6.82\times10^{14} $  &$0.55\pm0.05$ &$0.50\pm 0.05$ &$18.95\pm 0.13$ &1.72/13  &$ 0.47\pm 0.02$\\

\hline                                  
\end{tabular}
\end{table*}

\subsection{Optical polarimetry of PSR J1023+0038}\label{Sec_J1023}
With the aim of enhancing the signal to noise ratio (S/N) of the observation, we first of all summed together the five images in each optical band. Starting from the results of aperture photometry performed on the stars of the field, we applied the method described in Sec. \ref{Opt_pola} in order to evaluate the degree of LP, passing by the fit of the S parameter (eq. \ref{eq_S_par}) with the function reported in eq. \ref{fit_cos}. 
The results are tabulated in Tab. \ref{Tab_S_par_J1023}.


    \begin{table}
\caption{Results of the optical polarimetry performed on J1023.}             
\label{Tab_S_par_J1023}      
\centering                          
\begin{tabular}{c c c c}        
\hline\hline                 
 $B$  &$V$ &$R$ &$i$  \\    
\hline                        
\multicolumn{4}{|c|}{P ($\%  $)}\\ \hline
 $1.17\pm 0.51$&$1.09\pm 0.27$  &$0.90\pm 0.17$  &$0.55\pm 0.22$ \\
 \hline
\multicolumn{4}{|c|}{P ($3\sigma  $ upper limit)}\\ \hline
 $2.7 \%$& --  & --  &$1.21\%$ \\
\hline                                   
\multicolumn{4}{|c|}{Interstellar/instrumental LP ($3\sigma  $ upper limit)}\\ \hline
$1.23\%$& $0.63\%$  & $0.48\%$  &$0.54\%$ \\
\hline  
\end{tabular}
\end{table}

If we set at a $ 3\sigma $ level the minimum significance needed to define a detection of LP, we obtain evidences of polarised emission just in $ V $- and $ R $- bands, the first being slightly higher than the latter.
Since in $ B $- and $ i $- band our measurements are consistent with zero at the $2.3\sigma$ and $2.5\sigma  $ confidence level, respectively, we could just evaluate $ 3\sigma $ upper limits to the LP, which are reported in Tab. \ref{Tab_S_par_J1023}. Differently from what happens in the $ B $-band, the $ i $- band upper limit is particularly constraining, suggesting that the LP degree gets lower and lower while moving to longer wavelengths. The $ B $-band polarisation measurement has instead a high uncertainty due to the presence of the full moon, the radiation of which is polarised due to reflection, and affects any LP optical measurement, above all at higher frequencies (i.e. the $ B $--band).


We then tried to search for any possible phase--dependent variation of the optical LP of J1023. To this aim, each of the five datasets for each band has been analysed separately, so that we could gain in the four bands the trend of the polarisation degree with respect to the orbital phase, the latter being obtained starting from the ephemeris reported in \citet{Archibald2009}.

\begin{figure}
\centering
\includegraphics[scale=0.4]{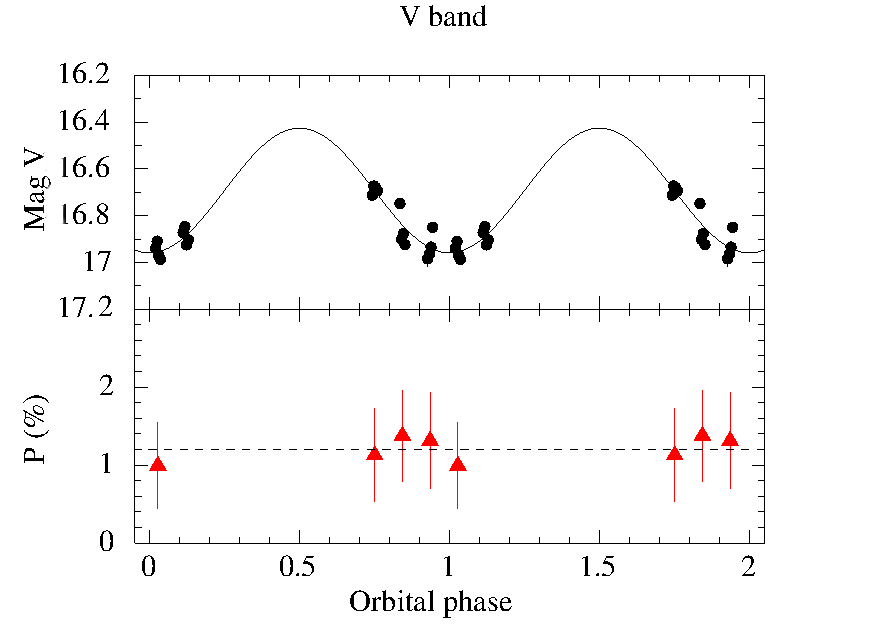}
\caption{\textit{Upper panel}: $ V $--band light curve of J1023 (magnitude vs. orbital phase). The solid line represents the fit of the light curve with a sinusoidal function with period fixed to 1 (corresponding to 4.75 hr; \citealt{Archibald2009}). \textit{Bottom panel}: $ R $--band polarisation curve represented as a function of the orbital phase of the system. The $ >1\sigma $ detections are represented with red triangles, whereas we indicated the $ 3\sigma $ upper limits with red arrows, where necessary. Superimposed, the fit of the $ >1\sigma $ detections with a constant function (dashed line; $ \chi^2/\rm dof  $=$0.28/3$). Errors are represented at the 68$\%$ confidence level. Two periods are drawn for clarity.}
\label{comparison_Vband}
\end{figure}

\begin{figure}
\centering
\includegraphics[scale=0.4]{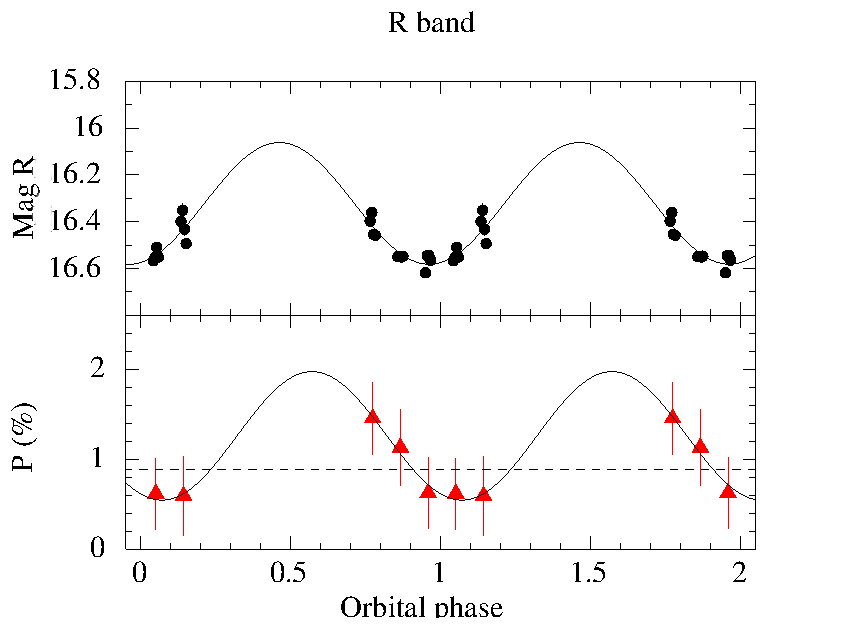}
\caption{\textit{Upper panel}: $ R $--band light curve of J1023 (magnitude vs. orbital phase). With a solid line, a fit of the light curve with a sinusoidal function with period fixed to 1 (corresponding to 4.75 hr; \citealt{Archibald2009}) is reported. The $ \chi^2/\rm dof  $ of the fit is 11.35/17. \textit{Bottom panel}: $ R $--band polarisation curve represented with respect to the orbital phase of the system. Superimposed, the fits with a constant function (dashed line; $ \chi^2/\rm dof  $=$3.66/4$) and with a sinusoidal function with period fixed to 1 (solid line; $ \chi^2/\rm dof $ =$0.11/2$). Errors are represented at the 68$\%$ confidence level. Two periods are drawn for clarity.}
\label{comparison_Rband}
\end{figure}

In the $ B $-- and $ i $--bands in some cases the S/Ns of the single observations were not high enough even to permit $ 1\sigma $ significance detections of LP. For this reason we could not detect any possible variations of LP with the orbital phase of the system.
In the $ R $-- and $ V $-- bands we could instead obtain $ > 1\sigma $ detections of LP in almost all datasets, in particular in the $ R $--band, where the significance of each measurement is always $ >2.3\sigma $, and exceeds $ 3\sigma $ in one case. While in the $ V $--band the polarisation trend seems to be almost constant with the orbital phase (Fig. \ref{comparison_Vband}, bottom panel; excluding the upper limits, a constant fit of the remaining points gives as a result $ \chi^2/\rm dof=0.3/3 $), in the $ R $--band the curve clearly deviates from being constant (the fit with a constant gives in particular $ \chi^2/\rm dof =7.32/4$). In particular, the shape of the polarisation curve seems to suggest a sinusoidal trend (Fig. \ref{comparison_Rband}, bottom panel). The fit with a sinusoidal function with period fixed to the orbital period of J1023 (4.75 hr) gives an acceptable $ \chi^2/\rm dof $ ($0.11/2$). Moreover, it is interesting to compare the $ R $--band light curve with the polarisation curve in this band. The positions of the maxima of the two curves are both located around phase 0.5 (superior conjunction of the companion star, i.e. when the observer sees the irradiated face of the donor), the two positions being consistent with each other within $ 1\sigma $ (maximum of the light curve phase: $0.46\pm 0.01$; maximum of the polarisation curve phase: $0.57 \pm 0.13$; errors at the $68\%$ confidence level).
It must be noted that the fit of the $ R $--band polarisation curve with a constant model is still preferable to the constant + sinusoidal one, despite the high $\chi^2/\rm dof$ of the former (according to an $F$-test, the addition of a
sinusoidal function gives a chance probability of $\sim 1.5 \times
10^{-2}$, corresponding only to a marginally 2.6$\sigma$ improvement in
the fit). Nevertheless, an intriguing increase of the polarisation degree towards phase 0.5 is clearly observed (Fig. \ref{comparison_Rband}).

\subsection{Optical polarimetry of XSS J12270-4859}
\begin{figure*}
\centering
\includegraphics[scale=0.4]{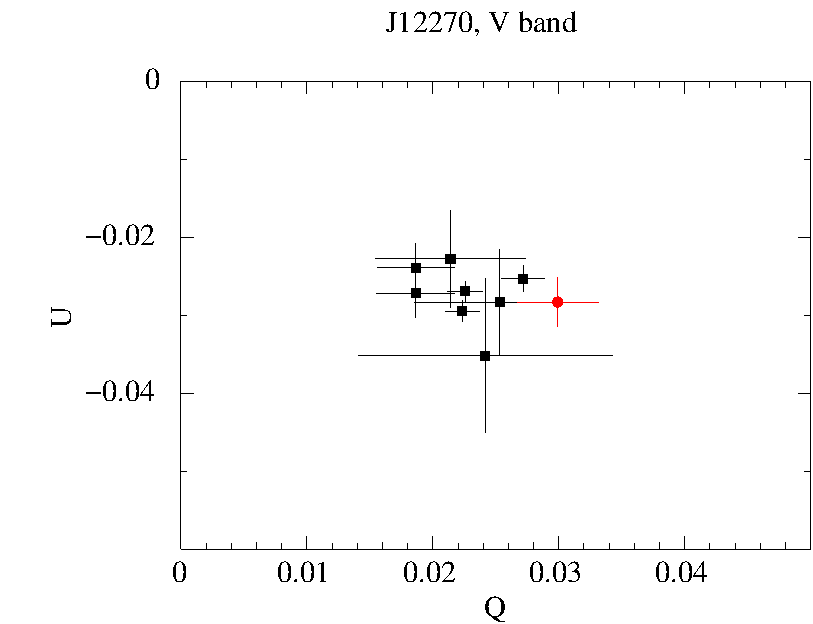}
\includegraphics[scale=0.4]{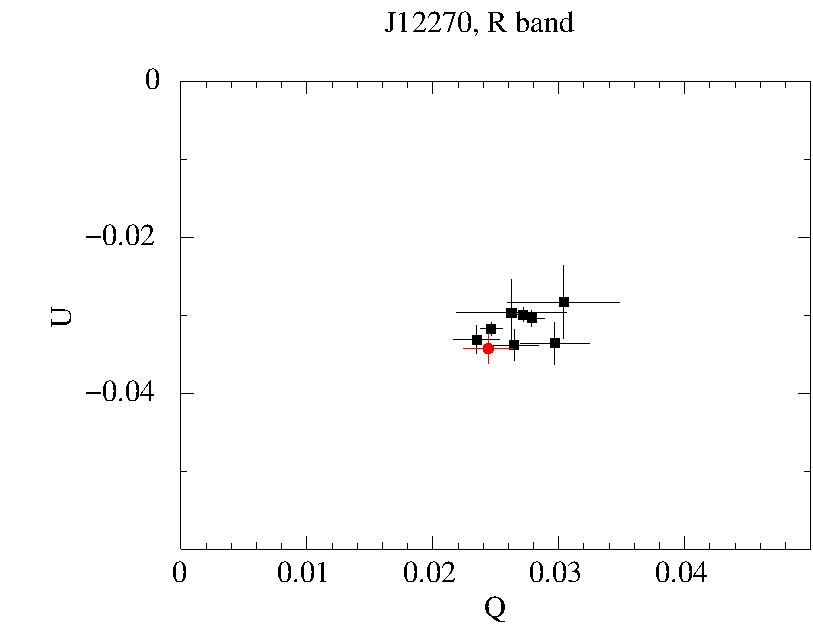}
\includegraphics[scale=0.4]{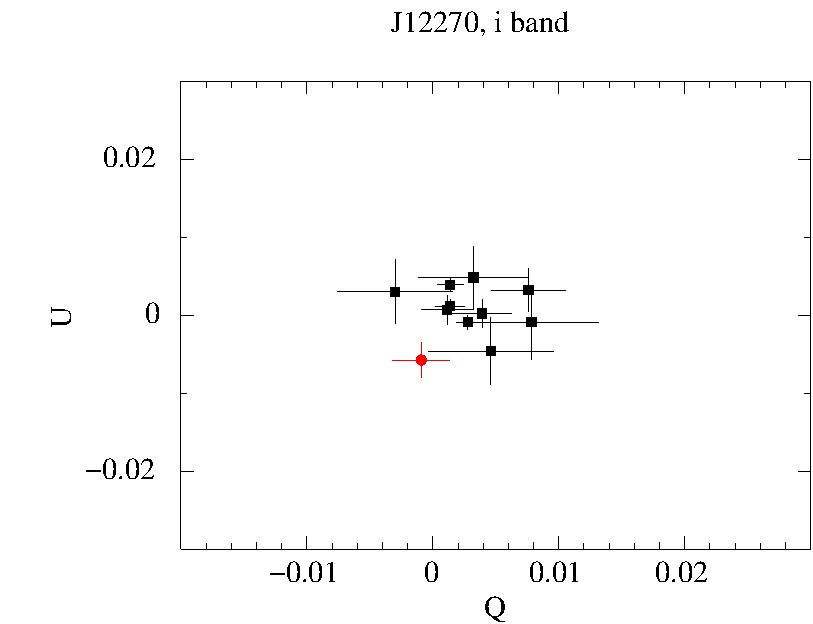}
\caption{From top to bottom, left to right: $U$ vs. $ Q $ for the averaged images in the optical $ VRi $ filters for J12270 (red dot) and a group of reference field stars (black squares). The reference stars and the target cluster around a common value in all bands, which testifies that J12270 is probably not intrinsically linearly polarised.}
\label{Q_U}
\end{figure*}

Following the same procedure performed on J1023, even for J12270 we averaged all the images collected in each optical band, in order to enhance the S/N and in this way to evaluate the average optical LP of the target. As first step we managed to estimate the Stokes parameters $ Q $ and $ U $ (eq. \ref{stokes}) of J12270 and of a group of comparison stars in the same field of the target, that were supposed to be unpolarised. The results are shown in Fig. \ref{Q_U} for the $ V $, $ R $ and $ i $--band, where the Stokes parameters are represented in the $ Q $--$ U $ plane for each filter. Due to the effect of the Moon, the Stokes parameters in the $ B $-- band were too dispersed to allow such study. As shown in the figure, the Stokes parameters of the reference stars cluster quite well around common values in each plot; these values represent the corrections that one has to apply in each band to the Stokes parameters of the target, in order to eliminate all the possible interstellar and instrumental contributions to the LP of its emitted radiation. This clustering testifies that the objects chosen as reference are indeed intrinsically not polarised (if they were, they would act independently of each other, and no clustering would be present). The fact that the target might show in the $ Q $--$ U $ plane a different behaviour with respect to the unpolarised reference stars could be a first hint of a possible intrinsic LP of its radiation. As clearly visible in Fig. \ref{Q_U}, this is not the case of J12270.

We then continued the analysis by evaluating the $ S $ parameter of J12270 at the different angles $\Phi$ (eq. \ref{eq_S_par}). Since the field of the target is sufficiently crowded, we could use the $o-$ and $e-$ fluxes of the group of reference stars in eq. \ref{eq_S_par}. The results of the fit of $ S(\Phi) $ with eq. \ref{eq_S_par} are tabulated in Tab. \ref{Tab_S_par_J12270}.

    \begin{table}
\caption{Results of the optical polarimetry performed on J12270. In particular, the LP degree corresponds to the semi-amplitude of the sinusoidal function used for the fit of the $ S $ parameter, reported in eq. \ref{eq_S_par}, whereas the residual interstellar and instrumental polarisation is indeed obtained as the constant parameter of the same fit.}         
\label{Tab_S_par_J12270}      
\centering                          
\begin{tabular}{c c c c}        
\hline\hline                 
 $B$  &$V$ &$R$ &$i$  \\    
\hline                        
\multicolumn{4}{|c|}{P ($\%  $)}\\ \hline
 $1.77\pm 1.33$&$0.73\pm 0.59$  &$0.40\pm 0.34$  &$0.74\pm 0.41$ \\
 \hline
\multicolumn{4}{|c|}{P ($3\sigma  $ upper limit)}\\ \hline
 $5.76 \%$& $2.50 \%$  & $1.42 \%$  &$1.97 \%$ \\
\hline                                   
\multicolumn{4}{|c|}{Interstellar/instrumental LP ($3\sigma  $ upper limit)}\\ \hline
$3.30\%$& $1.27\%$  & $0.93\%$  &$1.45\%$ \\
\hline  
\end{tabular}
\end{table}

As can be observed in Tab. \ref{Tab_S_par_J12270}, the significance of our findings is below $ 2\sigma $ in all bands. For this reason we could just evaluate $3\sigma$ upper limits to the optical LP of the target, that are reported in Tab. \ref{Tab_S_par_J12270} together with the upper limits to the possible residual interstellar and instrumental polarisation derived from the $S$ parameter fit with eq. \ref{fit_cos}. 

We tried to search for possible variations of the LP with the orbital phase of the system, being possible that in the process of averaging all the collected images at different orbital phases some cancellation effects came to pass.
However, in none of the analysed epochs a $ \geqslant 1\sigma $ polarisation detection was actually possible.  
\section{Discussion}
\subsection{Thomson scattering induced LP}
Unpolarised radiation emitted from an astrophysical source can be polarised due to different physical phenomena. The most frequent among them is electron scattering (Thomson scattering), 
according to which a free electron with a certain oscillation direction absorbs the incident, unpolarised radiation and re-emit it as linearly polarised in the same direction of its oscillation.
Thomson scattering in an astrophysical scenario can take place in different cases, the most common being the interaction of radiation with free electrons in the interstellar medium that can be found along the line of sight between the source of radiation and the observer. In this case, the induced polarisation has been demonstrated to have a precise wavelength dependency following the so--called Serkowski law \citep{Serkowski75}. In general, a rough estimate of the maximum expected interstellar contribution to the interstellar polarisation can be simply derived with the empirical formula \citep{Serkowski75}:
\begin{equation}\label{Serk}
P_{int, max}\leqslant 3 A_{V},
\end{equation}
where $ A_{V} $ is the absorption coefficient of the target in analysis in the optical $ V $--band. In the case of LMXBs, this contribution to the LP is usually treated as an effect to be corrected for, in order to search for other more interesting (and possibly intrinsic to the source) polarising phenomena. In particular, for these sources, and still in the field of polarisation induced by Thomson scattering, an intriguing possibility is that unpolarised light emitted from any system component (i.e. the companion star, the NS and even the accretion disc) may interact with the free electrons in the ionised accretion disc itself. In this case, as shown in \cite{Brown78} and \cite{Dolan84}, the resulting LP degree should not exceed a few per cent in the optical due to cancellation effects. Moreover, the LP degree should not depend on wavelength: the polarisation spectrum is expected to reproduce the spectrum of the incident radiation, since the Thomson cross section is known to be wavelength independent (the hot accretion disc being the principal responsible of the scattering in LMXBs, the level of LP is expected to increase with frequency in the optical regime). On the other hand, the polarisation degree due to Thomson scattering has a dependency on the scattering angle \citep{Serego}, and in the case of binary systems on the inclination $ i $ of the binary orbit \citep{Dolan84}. 
Interestingly, in case LP is induced by Thomson scattering, the Stokes parameters $ Q $ and $ U $ and the LP degree are expected to have a sinusoidal dependency on the orbital phase of the system (\citealt{Brown78}; \citealt{Dolan1988}; \citealt{Dolan89}).

In the case of J1023, all these considerations allow us to hypothesize that the measured LP (Tab. \ref{Tab_S_par_J1023}) may have its origin in the Thomson scattering of radiation with the free electrons in the accretion disc. 

\begin{figure}
\centering
\includegraphics[scale=0.105]{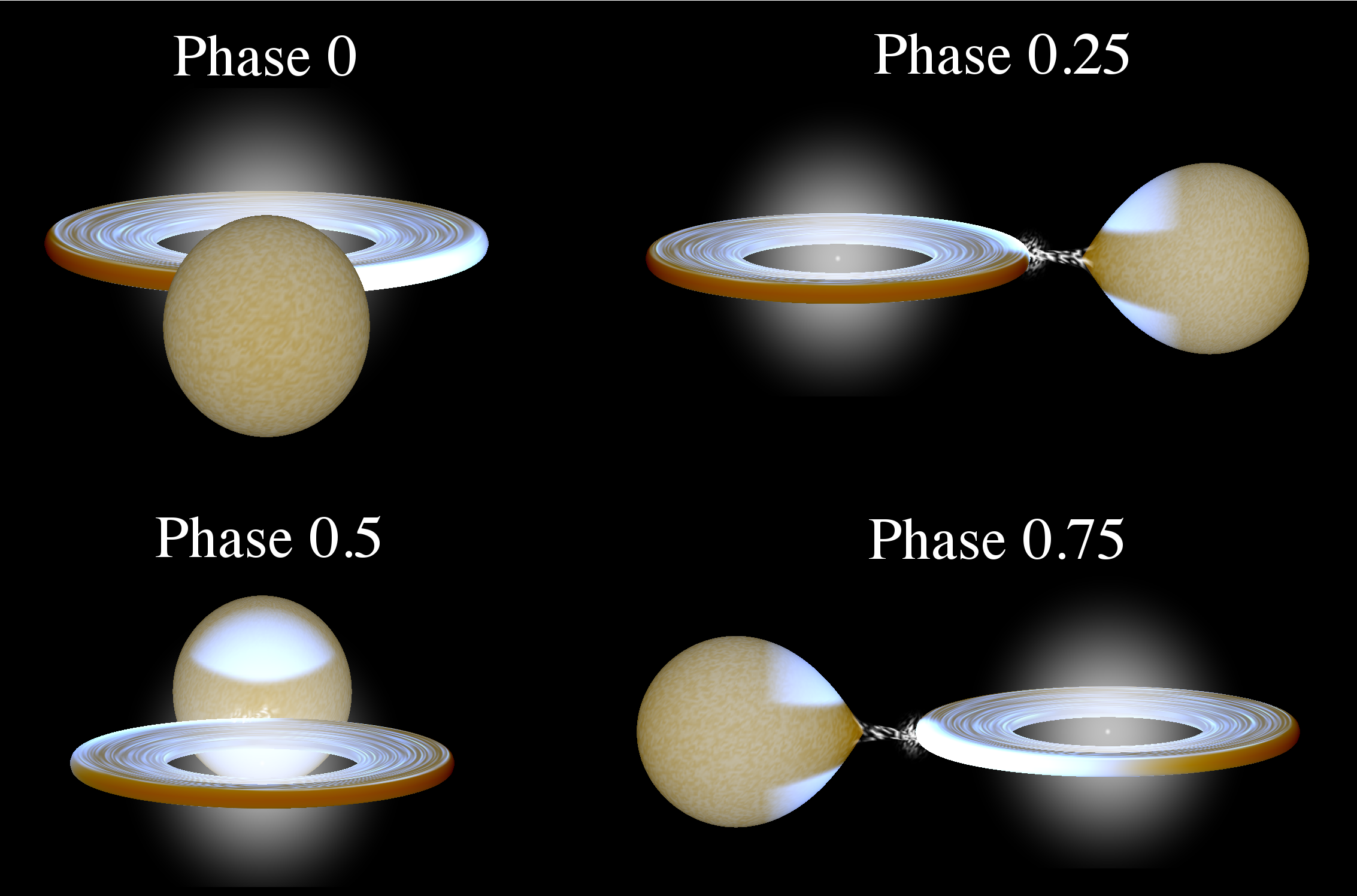}
\caption{Representation of a LMXB with the same orbital period of J1023 (4.75 hr) and a fictitious inclination $ i=80^\circ $, as seen from an external observer, obtained with the \textit{BinSim} software (available at http://www.phys.lsu.edu/$\sim$rih/binsim/). The four principal orbital phases of the system are taken into account.}
\label{Fig_all_phases}
\end{figure}

A possible evidence in favour of our hypothesis is the observation of the hinted tendency of the $ R $--band polarisation curve to oscillate at the system 4.75 hours orbital period (Fig. \ref{comparison_Rband}). In particular, it is interesting that the position of the LP maximum seems to coincide within the errors with that of the light curve, i.e. phase $ \sim $0.5 (i.e. superior conjunction -- when the observer sees the irradiated face of the companion star). In order to try to model qualitatively what we observed, let's consider for simplicity a LMXB with inclination $ i\sim 90^{\circ} $ (Fig. \ref{Fig_all_phases}). The optical radiation coming from such a system will be the combination of the emission of the companion star and that of the accretion disc, the X-ray emission of the internal disc being partially reprocessed by the external radii. 

When the companion star of the system is in phase of superior conjunction (phase 0.5, Fig. \ref{Fig_all_phases}), the accretion disc is found exactly along the line of sight between the observer and the companion star, causing the radiation emitted from both the companion star and the accretion disc to cross the accretion disc itself (if the latter is sufficiently extended and thick). In this way part of the radiation will interact with the free electrons that can be found in the disc and in the disc corona via Thomson scattering, thus resulting in a linearly polarised emission. 
When instead the system is at inferior conjunction (phase 0 -- when the observer sees the back side of the companion star), the optical radiation observed can be again ascribed to the same sources, even if the disc will be partially shaded by the companion star. However the configuration is now inverted, and the radiation emitted from the companion star is not intercepted by the accretion disc and the corona. This will cause the responsible of the emission of polarised radiation to be the Thomson scattering of the photons produced by the -- partially shaded -- accretion disc only with the free electrons in the disc itself and in the corona, thus implying a decrease in the LP of the total optical emitted light, as observed (Fig. \ref{comparison_Rband}). 
In the case of J1023 moreover a propeller scenario was proposed in order to account for the X--ray emission observed during the accretor phase (\citealt{PapittoTorres2015}; Campana et al. in preparation), according to which part of the accreting material may be halted at the NS magnetosphere. In this way, lots of free electrons may enclose the system, which will enhance the probability of Thomson scattering of the emitted photons (even if equally at all the orbital phases). 

Our simple, qualitative model could thus give an explanation to the variability of the $ R $--band polarisation curve shown in Fig. \ref{comparison_Rband}.  A good test bed for this hypothesis would be the observation of a similar, sinusoidal trend of the LP with the orbital phase in the other optical bands too, that was unfortunately not allowed by our data (as shown in Fig. \ref{comparison_Vband} for the $ V $--band), probably due to a too low S/N and, at higher frequencies, to the contaminant presence of the full moon. Moreover, once the oscillation is confirmed, a full coverage of the orbital period would help in constraining some physical and geometrical properties of the system (see e.g. \citealt{Gliozzi98}).

Differently to what happens for J1023, in the case of the LMXB J12270 we do not possess any LP $ 3\sigma $ over zero detection in the optical. This is however consistent with our hypothesis, according to which the LP of J1023 would be induced by Thomson scattering in the accretion disc of the system. In fact from the last works published on the transitional pulsar J1023 (e.g. \citealt{Bogdanov2015} and references therein) we know that the source is in an accretion--disc dominated state since 2013, which means that an accretion disc is indeed present in the system and can therefore be the cause of the observed LP. On the contrary, J12270 is lingering in a radio pulsar state since the end of 2012 \citep{deMartino2015}. In this condition, an accretion disc can not be formed, the accreting material being wiped out from the radio emission of the pulsar. Consequently the radiation emitted from J12270 is definitely not expected to be linearly polarised (or at least not because of Thomson scattering with electrons in an ionised accretion disc).

\subsection{Other possible causes: relativistic particles jets}
An intriguing, possible origin of the linear polarisation in LMXBs can be the production of relativistic particles jets. Such phenomena emit for instance synchrotron radiation, which is known to be intrinsically linearly polarised up to tens of per cent, depending on the level of ordering of the magnetic field at the base of the jet.
Jets have been observed both in BH and NS LMXBs, and in the case of BH LMXBs their coupling with the accretion state of the source has been extensively pointed out.
In particular, accretion states associated with hard X--ray spectra appear to be linked to the production of a relatively steady, continuously replenished and partially self--absorbed outflow \citep{Fender01}, while major outbursts are associated with more discrete ejection events. 
In the radio band the jets observed during the hard X–ray states of LMXBs have a flat spectrum. Above certain frequencies this flat spectral component breaks to an optically thin spectrum corresponding to the point at which the entire jet becomes transparent. There are evidences both from some BH and NS X--ray binaries that this break occurs around the NIR spectral region (e.g. \citealt{Migliari10}; \citealt{Gandhi11}). Thus the case is strong that there is a significant contribution of synchrotron emission in the NIR--optical spectral regimes of LMXBs, when a jet is emitted \citep{Russell11}.

For most BH and NS LMXBs in which jets have been detected, the optical and NIR polarisation do not exceed a few per cent; this means that the magnetic fields at the base of the jets are usually strongly tangled (e.g. \citealt{Schultz04}; \citealt{Russell11}; \citealt{Baglio_4U0614_2014}), the only known exception to this being the system Cyg X--1, for which on the contrary an ordered magnetic field has been measured \citep{Russell14}.

The X--ray luminosity of the two targets of this work, J1023 and J12270, is orders of magnitudes lower with respect to the NS LMXBs for which jets are usually inferred \citep{Deller2015}. However, \citet{Deller2015} found for J1023 a radio flux density and spectral index that were reminiscent of synchrotron emission due to a relativistic particles jet. For this reason it is not possible to completely rule out the possibility that the optical LP measured for J1023 (Tab. \ref{Tab_S_par_J1023}) could be due to a jet. All the upper limits and the detection of LP for J1023 (and for J12270, too) are in fact not constraining enough to rule out this scenario. If we hypothesize that a jet was emitted in both sources, we can try to estimate its maximum contribution to the total $ i $-- band flux (that is the optical band in which the jet contribution should be stronger), following what previously done for the quiescent LMXB Cen X--4 \citep{Baglio_cen2014}. In case of jet emission from a NS LMXB with tangled magnetic field one would expect to measure at most a polarisation degree of $\sim 5\%$ in the optical. If we consider our measured \textit{i}- band polarisation upper--limits (tabulated for the two sources in Tab. \ref{Tab_S_par_J1023} and \ref{Tab_S_par_J12270}) we thus obtain for the jet flux an upper limit of the $ (1.21\%/5)\cdot 100=24.2\% $ and the $ (1.97\%/5)\cdot 100=39.4\% $ of the total $ i $-- band flux for J1023 and J12270, respectively. These upper limits are not very constraining, meaning that the production of a jet from both systems can not be definitely excluded by our polarimetric observations.

We then sought some clues of the production of jets by looking at the spectral energy distributions (SED) of both sources, searching for any non--thermal (flat) component in the NIR that would distort the black--body shape of the typical LMXB infrared spectrum (see for example the infrared excess in the SED of the ultra--compact X--ray binary 4U 0614+091; \citealt{Migliari10}; \citealt{Baglio_4U0614_2014}). Starting from our photometry (Sec. \ref{opt_ima}), we could build the optical -- NIR SED of the two sources, that could in principle be sufficient to detect a possible jet and to observe the frequency of the jet break. Unfortunately, we did not possess contemporaneous datasets in all bands. However, in the case of J1023 our group possessed a more refined photometric analysis (already published in \citet{CotiZelati2014}) that referred to an earlier epoch (end of 2013), thanks to which we could obtain reliable parameters of the sinusoidal functions that describe the light curves trend in the optical. Under the hypothesis that the source did not vary considerably between 2013 and 2015, we could fix the semi--amplitudes of the 2015 light curves at the values reported in \citet{CotiZelati2014} for the filters that the two works have in common ($ i $ and $ R $) and we repeated the sinusoidal fit. In this way we gained the fluxes in the $ i $-- and $ R $-- bands at the same orbital phase as the NIR observations (we chose phase 0.59, that is the orbital phase at which the $ J $-- band data have been collected).

\begin{figure}
\centering
\includegraphics[scale=0.4]{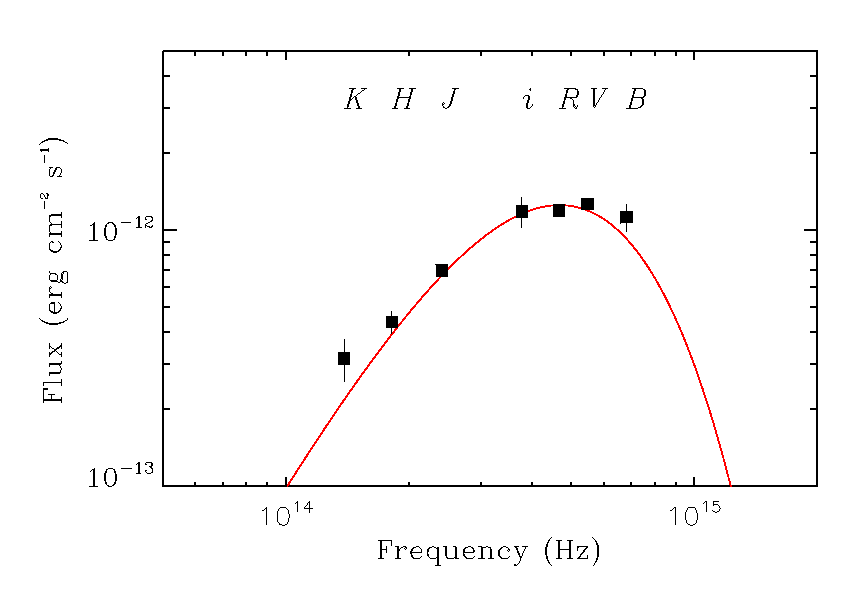}
\caption{NIR--optical averaged SED of J12270. Errors are indicated at the 68$\%$ confidence level. Superimposed, the fit of the SED with a black--body function.}
\label{NIR_SED_J12270}
\end{figure}

\begin{figure}
\centering
\includegraphics[scale=0.4]{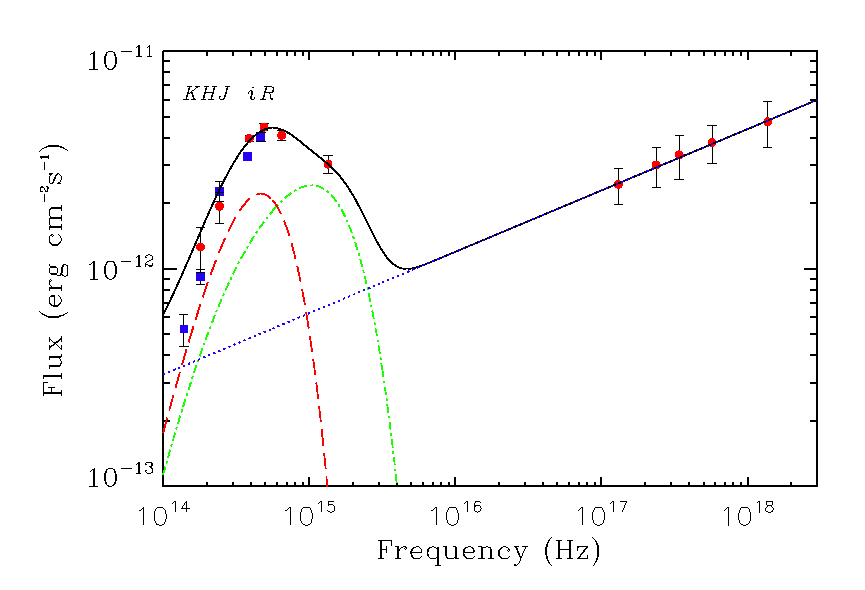}
\caption{NIR--optical contemporaneous SED of J1023 obtained in this work (orbital phase 0.59; blue squares); the NIR--X-rays SED of J1023 reported in \citet{CotiZelati2014} is also represented (red dots), with superimposed the fit with a model made by the companion star (red dashed line), the accretion disc (green dash-dotted line) and the shock emission powered by the NS spin down luminosity (blue dotted line).}
\label{comp_2014_2015}
\end{figure}

Concerning J12270 instead, we just built the averaged NIR--optical SED.
The SEDs of the two sources are shown in Fig. \ref{NIR_SED_J12270} and with blue dots in Fig. \ref{comp_2014_2015}.

Starting with the averaged SED of J12270, we first fixed the irradiation luminosity to the known value of the pulsar spin--down luminosity ($L_{\rm sd}=9\times10^{34}\rm erg\,s^{-1}$; \citealt{Roy2015}). With this prior, the SED can be qualitatively well described (despite the bad $ \chi^2/\rm dof=8.8/4 $, that is probably so high due to an underestimate of the errors) by a simple black body of an irradiated star, with free parameters the radius of the star ($R_{*}$) and the fraction $\eta$ of irradiation luminosity that heats the companion star. There is in particular no need to introduce any further non--thermal component in the infrared region of the SED, which makes the production of a jet in the system strongly unprobable. 
According to the best black--body fit, we found the fraction of irradiation luminosity that heats the donor star of the system to be $ \eta = 0.32\pm 0.05 $, whereas $R_{*}=0.33\pm 0.03\,R_{\odot}$ (errors are indicated at the 90$\%$ c.l.); the results do not change considerably until the surface temperature of the companion star remains fixed below 4000 K in the fit. 

Regarding J1023, we tried a comparison between our NIR--optical simultaneous points with the results reported in \citet{CotiZelati2014} (referring to observations carried out in Nov. 2013; Fig. \ref{comp_2014_2015}).  J1023 appears brighter in 2015 (blue squares) than in 2013 (red dots), but this could be simply due to the different orbital phases at which observations took place (around phase 0.9 in 2013 and around phase 0.6 in 2015, i.e. nearer to the maximum of the light curve). In order to reset this effect, we applied a rigid shift of $- 1.05\times10^{-12} \rm erg\,cm^{-2}\,s^{-1}$ to the 2015 points. The best fit of the 2013 SED represented in Fig. \ref{comp_2014_2015} seems to describe well also the optical points derived in this work, whereas our NIR points suggest a steeper decay towards lower frequencies (but the low frequency fit of the 2013 SED was not as well constrained as in 2015 by the dataset).
Interestingly however, the $ J-H $ colour measured in 2013 ($ 0.34\pm 0.31 $) is consistent with that derived in 2015 ($0.22\pm 0.05$), which means that the components involved in the 2014 NIR emission (the black body of the irradiated companion star + the accretion disc contribution) remained almost unchanged in 2015; in particular, there is no need to introduce the emission of a relativistic particles jet at the time of our observations to explain the observed SED.
Nevertheless, \citet{Deller2015} monitored J1023 over six months in 2014 (i.e. when the source was already in its accretion phase) at radio frequencies, observing rapidly variable but flat spectrum emission that possibly originates in an outflow from the system (or in a jet). In the epoch with most enhanced radio activity between those reported in \citet{Deller2015} (MJD 56674 -- 2014, Jan 17), the authors observed an average flux of $533 \pm 53 \, \mu\rm Jy$ in the 8-12 GHz frequency range, with spectral index $ \alpha=-0.27\pm0.07 $. We extended the radio flux back to the NIR, in order to compare it with our near--infrared results. What we observe however is that the radio predicts a lower flux than we observe, meaning that even in case of jet emission, the irradiated star (plus the accretion disc) emission is brighter than the jet itself. Therefore, it seems unlikely that the polarisation degree observed for J1023 in our work could be actually related to the emission of a jet.

We can thus conclude that the optical LP that we measure for J1023 is more probably due to Thomson scattering with the electrons in the accretion disc rather than to synchrotron emission from a jet, since no evidence for the presence of such phenomenon has been found in our optical and NIR photometry. 
 
\section{Conclusions}
In this work we present the results of the first optical ($ B $, $ V $, $ R $, $ i $--band) polarimetric study of the two transitional pulsars PSR J1023+0038 and XSS J12270-4859, based on observations carried out in 2015, February 8 and 9 with the ESO NTT equipped with EFOSC2. In addition, we also observed both systems in the NIR on 2015, February 7 with the NTT equipped with SOFI. Whereas for J12270 the degree of optical polarisation is consistent with 0 within 3$ \sigma $ in all bands, in the case of J1023 we found a low ($ \sim 1\% $; Tab. \ref{Tab_S_par_J1023}) but significant polarisation degree in the $ V$ and $ R $ bands. The level of optical polarisation has a slight increasing trend with frequency and, in the case of the $ R$ band, varies (possibly sinusoidally) with the orbital phase of the system, reaching the highest value around phase 0.5, i.e. superior conjunction (when the observer sees the irradiated face of the companion star). This peculiar behaviour is reminiscent of polarisation induced by Thomson scattering with the free electrons that can be found in the accretion disc of the system, that intercept the radiation emitted from all the components in the system depending on the orbital phase, thus causing the observed modulation of the polarisation degree (Fig. \ref{comparison_Rband}). Since we do not detect polarised light from J12270, this interpretation is also in accordance with the different states in which the two sources were lingering at the time of our observations (J1023: accretion; J12270: radio pulsar).

The measured polarisation degree of J1023 could in principle also be ascribed to the emission of synchrotron radiation from a relativistic particles jet. However the NIR-optical SED of J1023 is well described by the superposition of the two black-bodies of the companion star and the accretion disc (Fig. \ref{comp_2014_2015}), and no flat non-thermal component in the NIR that could hint at the emission of synchrotron radiation from the system is observed. This fact, together with the non-detection of a 3$ \sigma $ over zero polarisation degree in the $ i -$band (i.e. the one in which the jet should contribute the most, in the optical regime) let us retain the real emission of a jet at the time of our observations strongly unlikely, in favour of our first interpretation of the measured optical polarisation.
Follow up polarimetric optical observations covering the whole orbital period of the system are strongly encouraged in order to confirm the phase dependent oscillation of the LP.

Also in the case of J12270 the NIR--optical SED does not suggest the emission of relativistic particles jets, and can be fitted by a black--body function of an irradiated star with radius $ R_{*}=0.33\pm 0.03 \, R_{\odot} $ and albedo (i.e. the fraction of irradiation luminosity that heats the system donor star) $\eta=0.32\pm0.05$, withouth the need to introduce further components to the emission (as expected from the polarimetric analysis).
\begin{acknowledgements}
MCB thanks George Hau, Karla Aubel and the La Silla staff for their support during her observing run.
FCZ acknowledges the fruitful discussions with the international team on "The disk magnetosphere interaction around transitional ms pulsars" supported by ISSI (International Space Science Institute), Bern.
\end{acknowledgements}

\addcontentsline{toc}{chapter}{Bibliografia}


\end{document}